\documentclass[aps,prd,preprint,superscriptaddress,amsmath,amssymb,showpacs]{revtex4-1}
\usepackage{dcolumn}
\usepackage{graphicx}
\usepackage{float}
\usepackage{physics}
\usepackage[colorlinks=true,allcolors=blue]{hyperref}
\usepackage{mathrsfs}
\usepackage{inputenc}

\newcommand{\fb}{\mathfrak{b}}

\newcommand{\cA}{\cal A}

\newcommand{\nn}{\nonumber}
\newcommand{\be}{\begin{equation}}
\newcommand{\ee}{\end{equation}}
\newcommand{\bea}{\begin{eqnarray}}
\newcommand{\eea}{\end{eqnarray}}

\UseRawInputEncoding
\begin{document}
\title{Schwinger Effect in a Twice Anisotropic Holographic Model}

\author{Wen-Bin Chang}
\email{changwb@mails.ccnu.edu.cn}
\affiliation{ College of Intelligent Systems Science and Engineering, Hubei Minzu University, Enshi 445000,
People's Republic of China}

\date{\today}

\begin{abstract}
In this work, we investigate the Schwinger effect in a twice anisotropic holographic QCD model that incorporates both spatial and magnetic anisotropies.
Using the AdS/CFT correspondence, we calculate the total potential of a particle-antiparticle pair to evaluate how these anisotropies affect the holographic Schwinger effect.
Our calculations reveal that the magnetic field, characterized by parameters $c_B$ and $q_3$, consistently enhances the Schwinger effect by lowering and narrowing the potential barrier. In contrast, increasing the spatial anisotropy, parameterized by $\nu$, raises and widens the barrier, thereby suppressing the process.
These findings suggest the significance of treating both anisotropies concurrently for a realistic description of particle production in HIC.

\end{abstract}

\maketitle

\section{Introduction}\label{sec:01_intro}

In high-energy physics, heavy-ion collisions (HIC) conducted at facilities such as the Relativistic Heavy Ion Collider (RHIC) and the Large Hadron Collider (LHC) offer a unique opportunity to study physical phenomena under extreme conditions. 
These experiments are designed to create and investigate the quark-gluon plasma (QGP), a strongly coupled fluid composed of deconfined quarks and gluons\cite{SHURYAK198071,RevModPhys.81.1031}.
Within this extreme physical environment, one particularly compelling non-perturbative phenomenon is the Schwinger effect-the production of particle-antiparticle pairs from the quantum vacuum under a strong external electromagnetic field. The intense electromagnetic fields generated in HIC provide a promising environment for its observation, making the study of this effect crucial for understanding particle production mechanisms and the quantum structure of the vacuum.

AdS/CFT correspondence provides a natural framework for studying Schwinger pair production at strong coupling \cite{Maldacena:1997re,Gubser:1998bc,Witten:1998qj}. 
In the holographic context, the Schwinger effect was first formulated for $\mathcal{N} = 4$ super-Yang Mills (SYM) theory by Semenoff and Zarembo, and has since been extended to many gravitational backgrounds \cite{Semenoff:2011ng}.
The Schwinger effect in a confining phase was analyzed by Sato and Yoshida through a potential analysis of the corresponding holographic D3-brane and D4-brane backgrounds\cite{Sato:2013dwa}. 
The influence of higher-derivative Gauss-Bonnet terms on the holographic Schwinger effect was studied in a confining AdS soliton background \cite{Zhang:2015bha}.
A holographic calculation of the vacuum decay rate due to the Schwinger effect in confining gauge theories was presented in \cite{Hashimoto:2014yya}.
\cite{Zhang:2017egb} investigate the Schwinger effect in a strongly coupled plasma, focusing on the influence of a moving D3-brane and analyzing test particle pair production for motion both transverse and parallel to the plasma wind.
The influence of black hole spin on the potential barrier for Schwinger pair production was investigated holographically in the Myers-Perry spacetime in \cite{Cai:2023cjl}.
The combined effects of higher derivative corrections and a string cloud background on the holographic Schwinger effect were investigated across both confined and deconfined phases in \cite{Zhu:2024pdx}.
Relevant and insightful works can be found in \cite{Zhang:2020noe,Cai:2016jgr,Sonner:2013mba,Li:2018lsl,Lin:2024mct,Zhao:2024ipr,Chang:2024ksq,Zhou:2024fjj,Chen:2024ckb,Jiang:2024jxq,Chen:2024mmd,Chen:2023yug,Chen:2018vty,Wang:2025mmv,Chen:2019rez,Zhu:2025ucq,Chen:2020ath}
.

Off-central HIC create extreme early time conditions due to the rapid, oppositely directed motion of the colliding nuclei, which in turn produces strong transient magnetic fields ($10^{-1} \cdot  m_\pi^2\sim 15 \cdot  m_\pi^2$)\cite{Skokov:2009qp,Deng:2012pc}. 
Concurrently, experimental data from HIC indicate an early stage local anisotropy in the QGP, since the system expands predominantly along the collision axis and this anisotropy lasts for several fm/c \cite{Strickland:2013uga,Song:2020uvu}.   
The presence of these anisotropies in realistic HIC scenarios motivates investigations of the Schwinger effect within various anisotropic holographic models to capture these characteristics.
Previous studies have typically examined the influence of a single source of anisotropy, either spatial or magnetic, on the holographic Schwinger effect.
In \cite{Li:2022hka}, the authors employ a top-down holographic approach to systematically analyze the Schwinger effect and associated electric instabilities in the dual anisotropic gauge theory.
\cite{Zhou:2021nbp} investigates the holographic Schwinger effect in an anisotropic background and finds that anisotropy suppresses pair production compared to the isotropic case.
\cite{Zhu:2019igg} systematically analyzes the holographic Schwinger effect in the presence of weak and strong magnetic fields at RHIC and LHC energies and demonstrates that magnetic fields reduce the potential barrier to favor pair creation.
To simultaneously capture the spatial and magnetic anisotropies observed in realistic HIC, this work employs a five-dimensional Einstein-Maxwell-Dilaton model with three Maxwell fields.
The Schwinger effect is investigated within this twice anisotropic holographic model by calculating the total potential of a particle pair.

The paper is organized as follows. Sec.~\ref{sec:02} briefly reviews the anisotropic holographic model used in this study. Sec.~\ref{sec:03} presents the detailed potential analysis for the Schwinger effect and our numerical results. 
Finally, Sec.~\ref{sec:04} summarizes our findings and discusses potential extensions.
\section{ANISOTROPIC HOLOGRAPHIC MODEL}\label{sec:02}

In this section, we briefly review the anisotropic holographic QCD model introduced in \cite{Arefeva:2020vae,Arefeva:2022avn,Arefeva:2023jjh}, which has been used to study hot dense anisotropic QCD in the presence of an external magnetic field.
The 5-dimensional Einstein-Maxwell-dilaton gravity action with three Maxwell fields is given by
\begin{gather}
  {\ S } = \sqrt{-g} \left[ R 
    - \cfrac{f_0(\phi)}{4} \, F_0^2 
    - \cfrac{f_1(\phi)}{4} \, F_1^2
    - \cfrac{f_3(\phi)}{4} \, F_3^2
    - \cfrac{1}{2} \, \partial_{\mu} \phi \, \partial^{\mu} \phi
    - V(\phi) \right].
\end{gather}
In this model, the first ($F_0$), second ($F_1$), and third ($F_3$) Maxwell fields are respectively responsible for establishing a finite chemical potential, inducing spatial anisotropy, and providing an additional anisotropy sourced by the magnetic field.
Following \cite{Arefeva:2023jjh}, the holographic anisotropic background is described by a metric ansatz of the form
\begin{gather}
  ds^2 = \cfrac{L^2}{z^2} \ \fb(z) \left[
    - \, g(z) \, dt^2 + dx_1^2 
    + \left( \cfrac{z}{L} \right)^{2-\frac{2}{\nu}} dx_2^2
    + e^{c_B z^2} \left( \cfrac{z}{L} \right)^{2-\frac{2}{\nu}} dx_3^2
    + \cfrac{dz^2}{g(z)} \right] \! , \\
  \fb(z) = e^{2{\cA}(z)}, \nn
  \phi = \phi(z), \quad \,  \\
  \begin{split}
     \,\, A_0 &= A_t(z), \quad 
    A_{i}=0,\,\,i= 1,2,3,4, \\
     \quad
    F_1 &= q_1 \, dx^2 \wedge dx^3, \quad 
    F_3 = q_3 \, dx^1 \wedge dx^2\, . 
  \end{split}
\end{gather}
In this framework, $z$ is the holographic radial coordinate, $L$ is the AdS radius (set to one), $\fb(z)$ is the AdS deformation factor, $g(z)$ is the blackening function, $\nu$ is the anisotropy parameter characterizing primary spatial anisotropy, $c_B$ is the magnetic coefficient of secondary anisotropy induced by a magnetic field in the $x_3$ direction, and $q_1$ and $q_3$ are constant charges.

We adopt ${\cA}(z) = - \, cz^2/4 \, - (p - c_B q_3)
  z^4$, with the parameters $c = 4 R_{gg}/3$, $R_{gg} = 1.16$ and $p = 0.273$, to reproduce specific QCD phenomena \cite{He:2020fdi}.
In accordance with \cite{Arefeva:2023jjh}, solving the equations of motion yields the following expressions for the blackening function
\begin{gather}
  g(z) = e^{c_B z^2} \left[ 1 - \cfrac{\Tilde{I}_1(z)}{\Tilde{I}_1(z_h)}
    + \cfrac{\mu^2 \bigl(2 R_{gg} + c_B (q_3 - 1) \bigr)
      \Tilde{I}_2(z)}{L^2 \left(1 - e^{(2 R_{gg}+c_B(q_3-1))\frac{z_h^2}{2}}
      \right)^2} \left( 1 - \cfrac{\Tilde{I}_1(z)}{\Tilde{I}_1(z_h)} \,
      \cfrac{\Tilde{I}_2(z_h)}{\Tilde{I}_2(z)} \right)
  \right], \label{eq:4.42} \\
  \Tilde{I}_1(z) = \int_0^z
  e^{\left(2R_{gg}-3c_B\right)\frac{\xi^2}{2}+3 (p-c_B \, q_3) \xi^4}
  \xi^{1+\frac{2}{\nu}} \, d \xi, \qquad \ \label{eq:4.43-1} \\
  \Tilde{I}_2(z) = \int_0^z
  e^{\bigl(2R_{gg}+c_B\left(\frac{q_3}{2}-2\right)\bigr)\xi^2+3 (p-c_B
    \, q_3) \xi^4} \xi^{1+\frac{2}{\nu}} \, d \xi. \label{eq:4.43} 
\end{gather}
Accordingly, the Hawking temperature and entropy can be written as follows
\begin{gather}
  \begin{split}
    T &= \cfrac{|g'|}{4 \pi} \, \Biggl|_{z=z_h} = \left|
     - \, \cfrac{e^{(2R_{gg}-c_B)\frac{z_h^2}{2}+3 (p-c_B \, q_3)
         z_h^4} \, 
      z_h^{1+\frac{2}{\nu}}}{4 \pi \, \Tilde{I}_1(z_h)} \right. \times \\
    &\left. \times \left[ 
      1 - \cfrac{\mu^2 \bigl(2 R_{gg} + c_B (q_3 - 1) \bigr) 
        \left(e^{(2 R_{gg} + c_B (q_3 - 1))\frac{z_h^2}{2}}\Tilde{I}_1(z_h) -
          \Tilde{I}_2(z_h) \right)}{L^2 \left(1 
          - e^{(2R_{gg}+c_B(q_3-1))\frac{z_h^2}{2}}
        \right)^2} \right] \right|, 
    \\
    s = \ & \cfrac{1}{4} \left( \cfrac{L}{z_h} \right)^{1+\frac{2}{\nu}}
    e^{-(2R_{gg}-c_B)\frac{z_h^2}{2}-3 (p-c_B \, q_3) z_h^4},
  \end{split} \label{eq:4.48}
\end{gather}
where $z_h$ is the black hole horizon.

\section{SCHWINGER EFFECT IN ANISOTROPIC BACKGROUND}\label{sec:03}

Since the $x_1$ and $x_2$ directions can be considered simplified special cases of the $x_3$ direction when $c_B=0$ and $\nu=1$, for our calculations, we choose to align both the particle pair and the external electric field along the $x_3$ direction.
The coordinates are herein parameterized by 
\begin{equation}
\label{eq10}
\ t=\tau,\quad x_{3}=\sigma,\quad  x_{1}=x_{2}=0,\quad z=z(\sigma).
\end{equation}
Derived from the classical string action, the Nambu-Goto action is
\begin{eqnarray}
S= T_{F} \int d \sigma d \tau \mathcal{L}= T_{F} \int d \sigma d \tau \sqrt{-\operatorname{det} g_{\alpha \beta}},
\end{eqnarray}
where $T_{F}$ represents the string tension and $\operatorname{det} g_{\alpha \beta}$ is the determinant of the induced metric on the string worldsheet.
The Lagrangian density derived from the Nambu-Goto action is given as
\begin{eqnarray}
\mathcal{L}=\sqrt{\operatorname{det} g_{\alpha \beta}}= \frac{b(z)}{z^2}\sqrt{e^{c_B z^2}g(z)z^{2-\frac{2}{\nu}}+\dot{z}^2}.
\end{eqnarray}
Since the action is independent of the worldsheet coordinate $\sigma$, the associated Hamiltonian is conserved and given by
\begin{equation}
H=\mathcal{L}-\frac{\partial \mathcal{L}}{\partial \dot{z}} \dot{z}=\text { Constant }.
\end{equation}
Under the boundary condition
\begin{equation}
\label{eq14}
\ \dot{z}=\frac{dz}{d \sigma}=0, \quad z=z_{c} \ (z_{0}<z_{c}<z_{h}),
\end{equation}
wherein the probe D3-brane is located at the radial coordinate $z_{0}$, one obtains the differential equation
\begin{eqnarray}
\dot{z} =\frac{d z}{d \sigma}={\frac{\sqrt{e^{c_B z^2}g(z) z^{\frac{-4}{\nu}} \left(b(z)^2e^{c_B \left(z^2-{z_{c}}^2\right)}g(z) z_{c}^{2+\frac{2}{\nu}} -b(z_{c})^2 g(z_{c}) z^{2+\frac{2}{\nu}}\right)}}{b(z_{c}) \sqrt{g(z_{c})}}}.\label{zz}
\end{eqnarray}
Integration of Eq.~\ref{zz} directly provides the separation length $x$ for the test particle pair, which is expressed as 
\begin{equation}
\label{eq16}
\ x=2 \int^{z_c}_{z_0} dz{\frac{b(z_{c}) \sqrt{g(z_{c})}}{\sqrt{e^{c_B z^2}g(z) z^{\frac{-4}{\nu}} \left(b(z)^2e^{c_B \left(z^2-{z_{c}}^2\right)}g(z) z_{c}^{2+\frac{2}{\nu}} -b(z_{c})^2 g(z_{c}) z^{2+\frac{2}{\nu}}\right)}}} .
\end{equation}
Calculation of the combined Coulomb potential and static energy is performed according to
\begin{equation}
\label{eq17}
\ V_{(CP+SE)}= 2T_{F}\int^{z_c}_{z_0} dz \frac{
  b(z) b(z_c) \sqrt{g(z_c)} \sqrt{\frac{b(z)^2 e^{2 c_B z^2 - c_B z_c^2} g(z)^2 z^{-\frac{4}{\nu}} z_c^{2 + \frac{2}{\nu}}}{b(z_c)^2 g(z_c)}}
}{
  z^2 \sqrt{
    e^{c_B z^2} g(z) z^{-\frac{4}{\nu}} \left( -b(z_c)^2 g(z_c) z^{2 + \frac{2}{\nu}} + b(z)^2 e^{c_B (z^2 - z_c^2)} g(z) z_c^{2 + \frac{2}{\nu}} \right)
  }
}
.
\end{equation}
The critical electric field, $E_c$, is determined from the Dirac-Born-Infeld (DBI) action, given by 
\begin{equation}
S_{D B I}=-T_{D 3} \int d^4 x \sqrt{-\operatorname{det}\left(G_{\mu \nu}+\mathcal{F}_{\mu \nu}\right)}, \quad T_{D 3}=\frac{1}{g_s(2 \pi)^3 \alpha^{\prime^2}}.\label{dbix}
\end{equation}
Requiring the $S_{DBI}$ action to be free of singularities determines the critical field $E_c$
\begin{eqnarray}
E_{c} =T_{F} b(z_{0}) e^{\frac{c_B z_{0}^2}{2}} \sqrt{g(z_{0})} z_{0}^{-1 - \frac{1}{\nu}}.
\end{eqnarray}
By defining the dimensionless parameter $\alpha \equiv \frac{E}{E_{c}}$ , the total potential $V_{tot}$ of the pair can be formulated as
\begin{equation}
V_{t o t}=V_{(C P+S E)}-Ex.       
\end{equation}

\begin{figure}[H]
    \centering
      \setlength{\abovecaptionskip}{-0.1cm}
    \includegraphics[width=7cm]{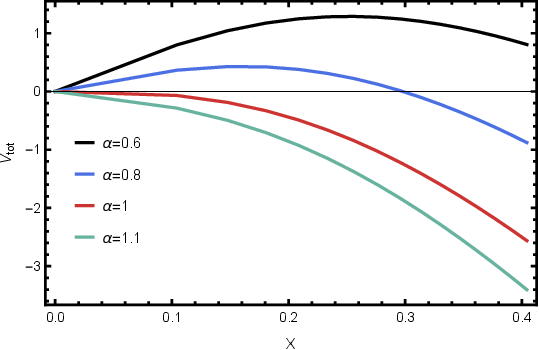}
    \caption{\label{fig1} The total potential $V_{tot}$ as a function of separation length $x$ for various electric field strengths at $c_B = - \, 0.1$, $\nu = 1.1$, $q_3 = 1$, $\mu = 0.1$ and $T= 0.6$.}
\end{figure}

\begin{figure}[H]
    \centering
      \setlength{\abovecaptionskip}{-0.1cm}
    \includegraphics[width=7cm]{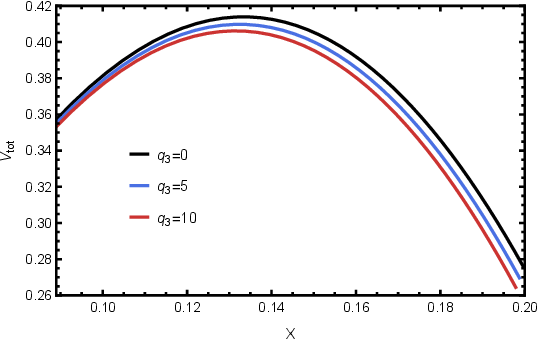}
    \caption{\label{fig2} The total potential $V_{tot}$ as a function of separation length $x$ for various magnetic charge $q_3$ at $c_B = - \, 0.1$, $\nu = 1.1$, $\mu = 0.1$ and $T= 0.6$.}
\end{figure}

\begin{figure}[H]
    \centering
      \setlength{\abovecaptionskip}{-0.1cm}
    \includegraphics[width=7cm]{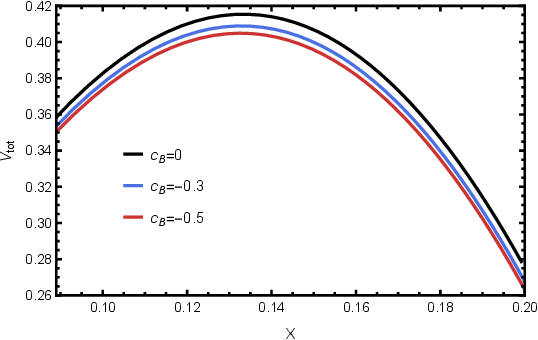}
    \caption{\label{fig3} The total potential $V_{tot}$ as a function of separation length $x$ for various magnetic coefficient $c_B$ at $\nu = 1.1$, $q_3 = 1$, $\mu = 0.1$ and $T= 0.6$.}
\end{figure}

Herein, we will investigate the Schwinger effect in the presence of two sources of anisotropy, one originating from spatial anisotropy and the other from an external magnetic field.
Considering the numerical results, we refer to Fig.~\ref{fig1}, which shows how the total potential $V_{tot}$ varies with separation length $x$ for various electric field strengths at $T= 0.6\ GeV$.
Observations indicate that the external electric field strength critically influences the rate of Schwinger pair production. The potential barrier diminishes with increasing field strength and disappears at the critical point ($\alpha = 1$). When $\alpha < 1$, pair production occurs via quantum tunneling through the barrier. For $\alpha >  1$, the suppression is removed and pair production becomes explosive, leading to vacuum instability.
The influence of the magnetic field on the Schwinger effect is characterized by evaluating the total potential $V_{tot}$ versus separation length $x$ over ranges of the magnetic coefficient $c_B$ and magnetic charge $q_3$.
Fig.~\ref{fig2} illustrates the impact of the magnetic charge $q_3$ on the potential barrier by plotting the total potential against the separation distance with $\alpha$ fixed at 0.8.
As shown in Fig.~\ref{fig2}, an increase in magnetic charge reduces both the height and width of the potential barrier, thereby increasing the tunneling probability of virtual particle–antiparticle pairs and consequently enhancing the Schwinger effect.
To reveal the effect of the magnetic coefficient, Fig.~\ref{fig3} presents the total potential versus separation distance, wherein we increase the absolute magnitude of the magnetic coefficient $c_B$ while maintaining a fixed magnetic charge.
As depicted in Fig.~\ref{fig3} , the height and width of the potential barrier diminish with an increasing absolute magnitude of the magnetic coefficient $c_B$ , which in turn facilitates the Schwinger effect.

Collectively, these findings underscore that the potential barrier is significantly influenced by both magnetic field parameters, namely the magnetic coefficient $c_B$ and the magnetic charge $q_3$.
As revealed in Fig.~\ref{fig2} and Fig.~\ref{fig3}, increasing the absolute values of the magnetic parameters $c_B$ and $q_3$ consistently lowers and narrows the potential barrier.
This directly translates to an enhanced Schwinger effect, indicating that stronger magnetic fields facilitate pair production.
These results from the potential analyses are in qualitative agreement with the results presented in \cite{Zhu:2019igg,Zhu:2021ucv}.

\begin{figure}[H]
    \centering
      \setlength{\abovecaptionskip}{-0.1cm}
    \includegraphics[width=7cm]{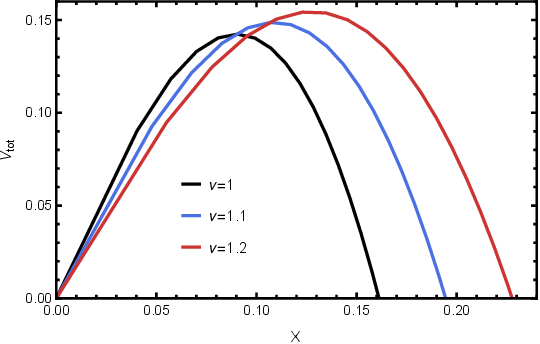}
    \caption{\label{fig4} The total potential $V_{tot}$ as a function of separation length $x$ for various anisotropy parameter $\nu$ at $c_B = - \, 0.1$, $q_3 = 1$, $\mu = 0.1$ and $T= 0.6$.}
\end{figure}

Turning to the influence of spatial anisotropy, parametrized by $\nu$, the potential barrier is also noticeably affected.
Fig.~\ref{fig4} examines the role of spatial anisotropy in the Schwinger effect by plotting the total potential $V_{tot}$ versus separation distance $x$ across varying anisotropy parameter $\nu$, with all magnetic parameters fixed.
In contrast to magnetic parameters that reduce the barrier, Fig.~\ref{fig4} indicates that, as $\nu$ increases, corresponding to greater spatial anisotropy, the potential barrier’s height and width both expand significantly.
This, in turn, suppresses the pair production process, thereby weakening the Schwinger effect, which is consistent with the calculations presented in \cite{Li:2022hka,Zhou:2021nbp,Chang:2023llr}.

\section{Conclusion and discussion}\label{sec:04}

In this work, we have investigated the Schwinger effect within an anisotropic holographic QCD model, focusing on the influence of two distinct sources of anisotropy: one arising from spatial anisotropy (parameterized by $\nu$) and the other from an external magnetic field (characterized by the magnetic coefficient $c_B$ and magnetic charge $q_3$).
By utilizing the AdS/CFT correspondence, we calculated the total potential of the particle pair and analyzed the behavior of the potential barrier that governs the pair production phenomenon under the influence of these anisotropic parameters.
We first summarize our numerical findings and their physical implications and then discuss potential extensions for future work.

The influence of the magnetic field parameters, $c_B$ and $q_3$, was found to consistently enhance the Schwinger effect. 
As demonstrated in Fig.~\ref{fig2} and Fig.~\ref{fig3}, increasing the magnetic charge $q_3$ or the absolute magnitude of the magnetic coefficient $c_B$ leads to a reduction in both the height and width of the potential barrier, thereby promoting the Schwinger effect.
These findings are in qualitative agreement with previous studies \cite{Zhu:2019igg,Zhu:2021ucv} and suggest that stronger magnetic environments facilitate pair production.
In contrast to the magnetic case, increasing the spatial anisotropy, characterized by the parameter $\nu$, results in a significant increase in both the height and width of the potential barrier and consequently weakens the Schwinger effect.
The spatial anisotropy increases the energy required to separate a virtual pair, thereby causing a suppression effect consistent with previous findings \cite{Li:2022hka,Zhou:2021nbp,Chang:2023llr}.

In realistic HIC, both the magnetic field and the spatial anisotropy evolve rapidly in time. 
A natural extension would be to generalize our analysis to a time-dependent holographic background that incorporates the evolution of both of these effects.

\section*{Acknowledgments}


Wen-Bin Chang is supported by the Ph.D. Research Startup Project at Hubei Minzu University (Project No. RZ2500000857).

\bibliography{ref}
\end{document}